\documentclass[a4paper,11pt]{article}
\usepackage{pos}

\title{The two-dimensional and three-dimensional relations in the plateau emission in multi-wavelengths}
\ShortTitle{GRBs for high-$z$ cosmology}

\author*[a,b,c,d,e]{Maria Giovanna Dainotti}
\author[f,g]{Biagio De Simone}

\affiliation[a]{Division of Science, National Astronomical Observatory of Japan, 2-21-1 Osawa, Mitaka, Tokyo 181-8588, Japan}
\affiliation[b]{The Graduate University for Advanced Studies (SOKENDAI), Shonankokusaimura, Hayama, Miura District, Kanagawa 240-0115}
\affiliation[c]{Space Science Institute, 4765 Walnut St Ste B, Boulder, CO 80301, USA}
\affiliation[d]{Nevada Center for Astrophysics, University of Nevada, 4505 Maryland Parkway, Las Vegas, NV 89154, USA}
\affiliation[e]{Bay Environmental Institute, P.O. Box 25 Moffett Field, CA, California}
\affiliation[f]{Dipartimento di Fisica, Universit\'a di Salerno, Via Giovanni Paolo II, 132 I-84084 Fisciano (SA), Italy}
\affiliation[g]{INFN, Sezione di Napoli, Gruppo collegato di Salerno, Italy}

\emailAdd{maria.dainotti@nao.ac.jp}
\emailAdd{bdesimone@unisa.it}

\abstract{\textbf{Abstract}\\
Gamma-Ray Bursts (GRBs) are interesting objects for testing the emission models in highly energetic regimes and are very promising standardizable candles, given their observability at high redshift (up to $z=9.4$) that allows the extension of the Hubble diagram much further the limit of Supernovae Ia (SNe Ia), the most distant one being at $z=2.26$. In this study, we demonstrate that the fundamental plane relation involving the prompt peak luminosity in X-rays, the X-rays plateau-end luminosity, and the plateau-end rest-frame time is not only a robust benchmark for testing GRB emission models like the magnetar but also a promising avenue for high-$z$ cosmology exploration. First, we discuss the connection between the magnetar model and the GRB afterglow correlations. 
Second, through the simulation of GRBs, we count how many years are needed to achieve the same precision of modern SNe Ia samples in the estimation of $\Omega_{M}$.}

\FullConference{%
  Multifrequency Behaviour of High Energy Cosmic Sources - XIV\\
  12-17 June 2023\\
  Mondello, Palermo, Italy
}

\begin{document}
\maketitle

\section{Introduction}
GRBs are very energetic flashes of cosmological origin photons distributed isotropically in the sky, ranging from $\gamma$-ray and X-rays in the prompt emission to the X-rays, optical, and sometimes radio in the afterglow phase. They have been observed up to redshift $z=9.4$ \citep{Cucchiara2011}, much further than the SNe Ia at $z=2.26$ \citep{Rodney2016} and quasars (QSO, \citep{Colgain2022,Dainotti2022QSO,Dainotti2023QSO}) with the furthest being at $z=7.642$ \citep{Wang2021}. In the 42 $\%$ of GRBs observed by the Swift-XRT telescope (Swift mission, 2004-ongoing), it is possible to observe the plateau phase, namely, a flattening of the light curve after the prompt emission and before the late afterglow of the light curve. This feature is of significant astrophysical interest, given that it can be powered by magnetars. However, it's also an important instrument to achieve the standardization of GRBs since the afterglow properties have proven to be more regular than the prompt ones from a morphological point of view. GRBs require unbiased and intrinsic correlations among their parameters to be used as cosmological tools, and among all the proposed correlations for GRB standardization, an outstanding candidate is the fundamental plane relation, also called the 3D Dainotti relation, among the X-ray at the end of plateau emission ($L_X$), the plateau emission end-time in the rest-frame ($T^*_X$), and the 1-second peak prompt X-ray luminosity ($L_{p,_X}$) \citep{Dainotti2016,Dainotti2017,Dainotti2020,Dainotti2022ApJS}. This relation combines the 2D Dainotti relation ($L_X-T^*_X$) \citep{Dainotti2008,Dainotti2013,Dainotti2020,Dainotti2022ApJS,Dainotti2022MNRAS,Dainotti2022PASJ} and $L_X-L_{p,_X}$ \citep{Dainotti2011,Dainotti2015}. Notably, it exhibits the smallest intrinsic scatter ($\sigma_i=0.18\pm0.07$) among multidimensional correlations involving GRB plateaus \citep{DainottiLenartChraya}. 
The two-dimensional relation has also been discovered in optical \citep{Dainotti2020} and radio \citep{Levine}. In addition, \cite{Dainotti2022ApJS} discovered that, similarly to the X-ray emission, it is possible to extend the optical relation in three dimensions. It is important to note that these relations show compatibility in their parameters. 

Crucially, the fundamental plane fitting parameters remain independent of cosmological models, confirming the reliability of GRBs as cosmological probes. The fundamental plane relation is expressed as follows:

\begin{equation}
    \log_{10}L_X = \alpha \log_{10}T^{*}_X + \beta \log_{10}L_{p,_X} + \gamma,
    \label{eq:fundamentalplane}
\end{equation}

where $\alpha,\beta$ are the plane parameters and $\gamma$ is the normalization of the plane. This correlation can be corrected for selection biases and redshift evolution effects through the Efron and Petrosian (EP, \citep{EP1992}) method formulation. After applying the EP method, the fundamental plane relation expressed in Equation \ref{eq:fundamentalplane} becomes:

\begin{multline}
    \log_{10}L_X - k_{L_X}\log_{10}(1+z) = \alpha_{ev} (\log_{10}T^{*}_X - k_{T^{*}_X}\log_{10}(1+z)) \\ + \beta_{ev} (\log_{10}L_{p,_X} - k_{L_{p,_X}}\log_{10}(1+z)) + \gamma_{ev},
    \label{eq:fundamentalplaneEP}
\end{multline}

where $k_{L_X},k_{T^{*}_X},k_{L_p}$ are the evolutionary coefficients computed with the EP method that allow to correct the parameters $L_X,T^{*}_X,L_{p,_X}$ for selection biases or evolution effects with $z$. 

The current proceeding is organized as follows. 
First, in Section \ref{sec:magnetar}, we introduce the magnetar model and its connection with the plateau emission of GRBs.
%Second, in Section \ref{sec:GRBscombined}, we apply the fundamental plane relation with the current number of observed GRBs together with SNe Ia, Baryon Acoustic Oscillations (BAOs), and QSO, showing the significant contribution of GRBs that imply consistent results with the other probes in the estimation of the cosmological parameters but extending the Hubble diagram to further redshifts.
Second, in Section \ref{sec:GRBsalone}, we estimate the required number of GRBs with plateau emission to reach the same precision as SNe Ia in constraining the $\Omega_{M}$ parameter. 

\section{The magnetar model for GRBs and the plateau emission}\label{sec:magnetar}
GRBs are traditionally classified through a bimodal scheme \citep{Kouveliotou1993}: (1) Long GRBs (LGRBs), with a prompt duration $T_{90}>2\,sec$, a soft spectrum, and are associated with the core-collapse of massive stars; (2) Short GRBs (SGRBs), having $T_{90}<2\,sec$ and a harder spectrum, generated by the merging of compact objects, such as two neutron stars (NS-NS) or a NS with a black hole (BH).
The situation becomes more complicated when more subclasses are included, such as the Short GRBs with extended emission (SEE) that show a hard spectrum typical of SGRBs, but with a duration $T_{90}>2\,sec$. The existence of these events challenges the traditional LGRB-SGRB division. Consequently, a more robust classification scheme necessitates identifying common features among different GRB classes, and one such characteristic is the plateau emission.
The mechanism that powers the plateau can be explained by the fallback of materials onto a newly formed BH or the spin-down emission of an ultra-magnetized newborn millisecond NS, the so-called \textit{magnetar}. The debate about the most reliable model is still ongoing in the literature. However, the magnetar model has captured further attention since a magnetar can be either formed in the case of massive stars that undergo core-collapse or in the case of NS binaries merging \citep{Bernardini2015}: this implies that the magnetar model is a valid interpretation for the GRB emission both in the cases of LGRBs and SGRBs, thus paving the way to overcome the traditional LGRB-SGRB classification scheme.
In the magnetar scenario, the rotational energy is released very quickly in the first hours through the spin-down of the magnetic dipole, which naturally implies the presence of a long-lived central engine. In \citep{Rownlinson2014}, the authors show how the 2D Dainotti relation $L_X-T^{*}_X$ \citep{Dainotti2008,Dainotti2013,Dainotti2020,Dainotti2022ApJS,Dainotti2022MNRAS,Dainotti2022PASJ} can be explained by the magnetar central engine. This correlation can be written in the following form:

\begin{equation}
    \log_{10}L_X=a+b\log_{10}T^{*}_X
    \label{eq:2Drelation}
\end{equation}

where $T^{*}_X$ is the rest-frame plateau-end time (in $sec$) and $L_X$ is the plateau-end luminosity (in $erg/sec$). In linear form, this correlation can be written as $L_X=aT^{*}{b}_X$.
The newly formed magnetar can release its energy by dipole radiation, and, for simplicity, it is possible to ignore the transfer mechanism from the magnetar to the observed emission. There is an intrinsic association between the bolometric plateau-end luminosity and duration and, according to \citep{ZhangMeszaros2001}, these can be written as:

\begin{equation}
\begin{aligned}
    L_{plat} \sim (B^{2}P^{-4}R^{6})\\
    T_{plat} \sim 2.05(IB^{-2}P^{2}R^{-6}),
\end{aligned}
\label{eq:LTmagnetar}
\end{equation}

where $L_{plat}$ and $T_{plat}$ are the theoretical plateau-end luminosity and time, in $10^{49}erg/sec$ and $10^{3}sec$, respectively, $I$ is the moment of inertia ($10^{45}g\,cm^{2}$), $B$ is the magnetic field intensity ($10^{15}G$), $R$ is the neutron star radius ($10^{6}cm$), and $P$ is the initial rotational period ($10^{-3}sec$). Through Equations \ref{eq:LTmagnetar} a $L_{plat}-T_{plat}$ relation can be written as follows:

\begin{equation}
    \log_{10}(L_{plat})\sim \log_{10}(10^{52}I^{-1}P^{-2})-\log_{10}(T_{plat}).
    \label{eq:2Dmagnetar}
\end{equation}

Therefore, a correlation exists between $L_{plat}$ and $T_{plat}$ that can be roughly represented as $L_{plat} \propto T^{-1}_{plat}$, denoted as $b=-1$ based on predictions from Equation \ref{eq:2Drelation}. In a study by \citep{Dainotti2013}, we derived the parameter value $b=-1.07^{+0.09}_{-0.14}$ by fitting the 2D Dainotti relation to 101 Swift-XRT GRBs featuring plateau emission. This dataset spans observations from January 2005 to May 2011, with the Dainotti relation parameters adjusted for astrophysical biases using the EP method.
The expected normalization of the relation, denoted by the parameter $a$ and anticipated to be $52$ according to the magnetar model, aligns well with the observed value $a=52.73\pm0.52$ from the normalization given by the bolometric rest-frame energy band $1-10000\,keV$ for Swift-XRT. This outcome strongly suggests that the magnetar model is a highly reliable framework for explaining the plateau emission observed in the early afterglow of GRBs.

Another step forward in the connection of the $L_X-T^{*}_T$ and $L_{p,_X}-L_X-T^{*}_X$ with the magnetar scenario has been done in \citep{Stratta2018}. In the formulation of this work, the NS spinning down is capable of releasing a luminosity $L_{s}=I\Omega\dot{\Omega}$, where $\Omega=2\pi\nu$ is the spin rate and $I$ the moment of inertia. For NSs, it holds the relation $\dot{\Omega} \propto \Omega^{n}$, where $n=3$ is the braking index for the ideal magnetohydrodynamical conditions, while $n\leq 3$ is considered in the non-ideal case.

In the ideal scenario, the spin-down luminosity relation is the following:

\begin{equation}
    L_s=\frac{\mu^2}{c^3}\Omega^{4}(1+\sin^{2}{\theta}),
    \label{eq:idealspindown}
\end{equation}

where $\mu$ is the magnetic dipole moment and $\mu \propto B$, $B$ being the magnetic field, and $\theta$ the angle between the rotation axes and the magnetic field direction. This expression can be generalized to the non-ideal case, writing it as

\begin{equation}
    L^{n.i.}_s=L_s\big(\frac{\Omega}{\Omega_i}\big)^{-2\alpha},
    \label{eq:nonidealspindown}
\end{equation}

where $\Omega_i$ is the initial value, $n.i.$ means non-ideal, and $n=3-2\alpha$ considering $0<\alpha<1$. In \citep{DallOsso2011}, the energy budget in the relativistic external shock is given by the balance of injected energy from the magnetar spinning down and the loss of radiative energy according to

\begin{equation}
    \frac{dE}{dT}=L_s(T)-\epsilon\frac{E}{T},
    \label{eq:energyexternalshock}
\end{equation}

where $T$ as the observer frame time and the coefficient $\epsilon$ depends on the electron energy fraction and the shock evolution with time. For simplicity, the $\epsilon$ has been considered as a constant.

In \citep{Stratta2018}, a sample of 40 GRBs with plateau emission observed by Swift-XRT has been fitted through the phenomenological Willingale model \citep{W07} to estimate the properties $L_X,T^{*}_X,L_{p,_X}$. The sample, called golden sample, was selected according to the following criteria also adopted in \citep{Dainotti2016}:

\begin{itemize}
      \item The presence of a minimum 5 data points in the plateau;
    \item The steepness of the plateau $<41^{\circ}$;
    \item The fit should follow the Avni 1978 prescription \citep{Avni1978} for the fit.
\end{itemize}

\begin{figure}
    \centering
    \includegraphics[scale=0.35]{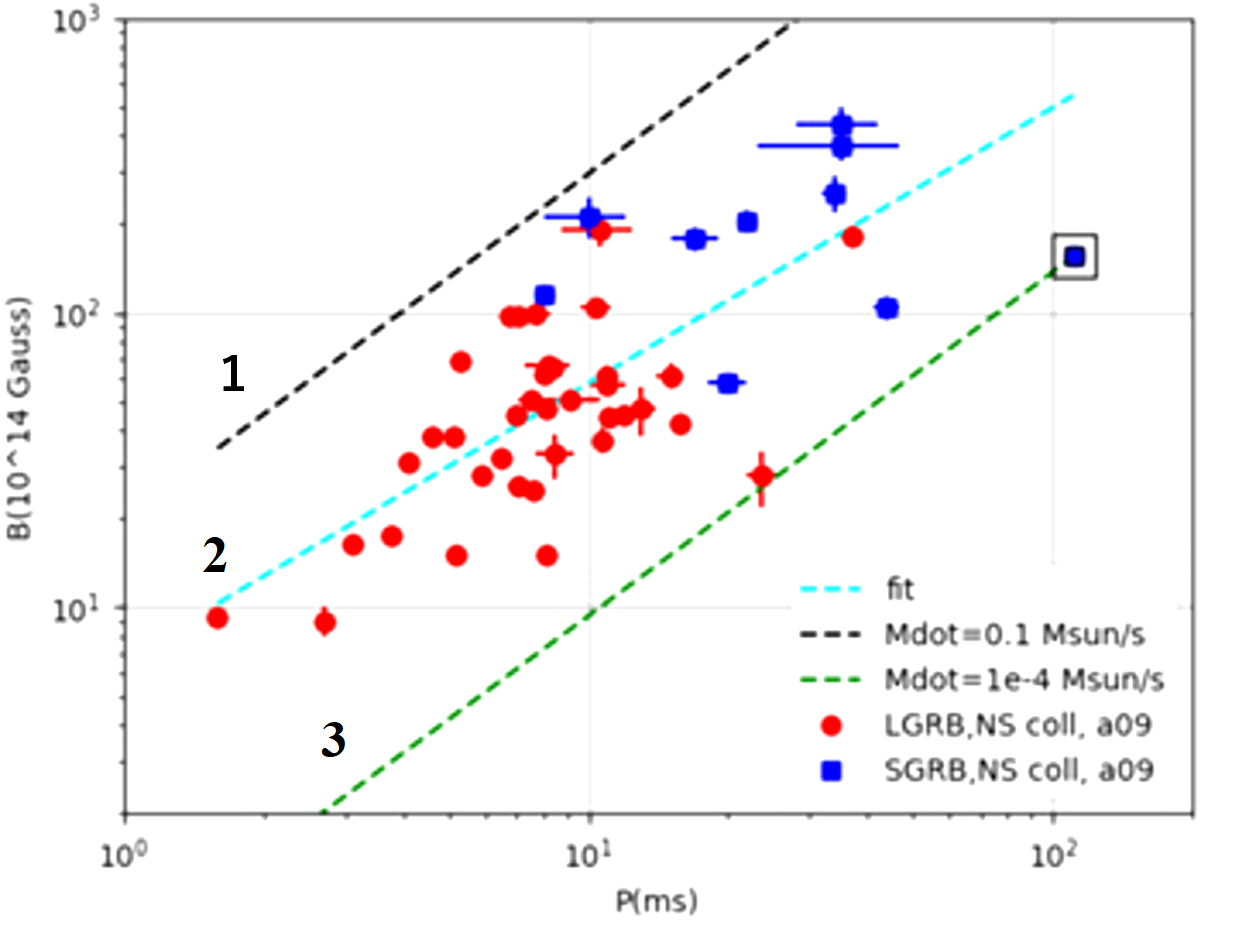}
    \caption{The $P-B$ diagram from \citep{Stratta2018}. The lines mark indicate the expected $P-B$ relations from accreting NSs with an accretion rate of $0.1$ (line 1) and $10^{-4}M_{\odot}/sec$ (line 3); line 2 indicates the best-fit relation.}
    \label{fig:PBrelation}
\end{figure}

This sample of GRBs, containing LGRBs and SEE, leads to the following fitting of the spin period-magnetic field ($P-B$) relation:

\begin{equation}
    \log_{10}B=(0.83\pm0.17)P + 0.84,
    \label{eq:PBrelation}
\end{equation}

where the uncertainty on the slope has been symmetrized. The fitting results are depicted as line 2 in Figure \ref{fig:PBrelation}. The $P-B$ correlation is anticipated based on the physics of spin-up lines observed in accreting neutron stars (NS) within Galactic binary systems. In this dataset, all data points fall between two boundary lines determined by the mass accretion rate, $\dot{M}$: $10^{-4}M_{\odot}/sec<\dot{M}<0.1M_{\odot}/sec$. These boundaries are illustrated as lines 1 and 3 in Figure \ref{fig:PBrelation}. Notably, these values align with observations in the prompt emission of gamma-ray bursts (GRBs), as $\dot{M}$ can be considered a proxy for $L_{p,_X}$.

The $B-P$ relation potentially indicates a connection to the $L_X-T^{*}_X$ relation since both links stem from magnetar properties. Furthermore, the inclusion of $L_{p,_X}$ in the fundamental plane relation, expressed in Equation \ref{eq:fundamentalplane}, alongside the plateau properties, facilitates linking prompt luminosity to plateau luminosity and time within the magnetar model framework.

An additional key insight from this analysis is the distinct clustering of SEE and LGRBs in different regions of the $P-B$ diagram. SEE events tend to cluster in the longer-duration end of the diagram, while LGRBs are concentrated in the region characterized by shorter periods and lower $B$ intensities, as illustrated in Figure \ref{fig:PBrelation}. This observation suggests that the fundamental plane relation serves as a discriminant tool among various GRB classes. This conclusion aligns with findings in \citep{DainottiPlatinum}, where the different classes of LGRBs associated with supernovae (SGRB-SNe), GRBs associated with kilonovae (SGRB-KNe), and the standalone SGRBs exhibit statistically distinct distances from the fundamental plane fitting for each class.

The fundamental plane relation has proven to be a reliable astrophysical instrument for discriminating among different GRB classes and GRB emission mechanism models, but it also has important implications in cosmology, as summarized in the next Sections.

\section{The application of GRBs as standalone cosmological probes}\label{sec:GRBsalone}
\begin{figure}
    \centering
    \includegraphics[scale=0.35]{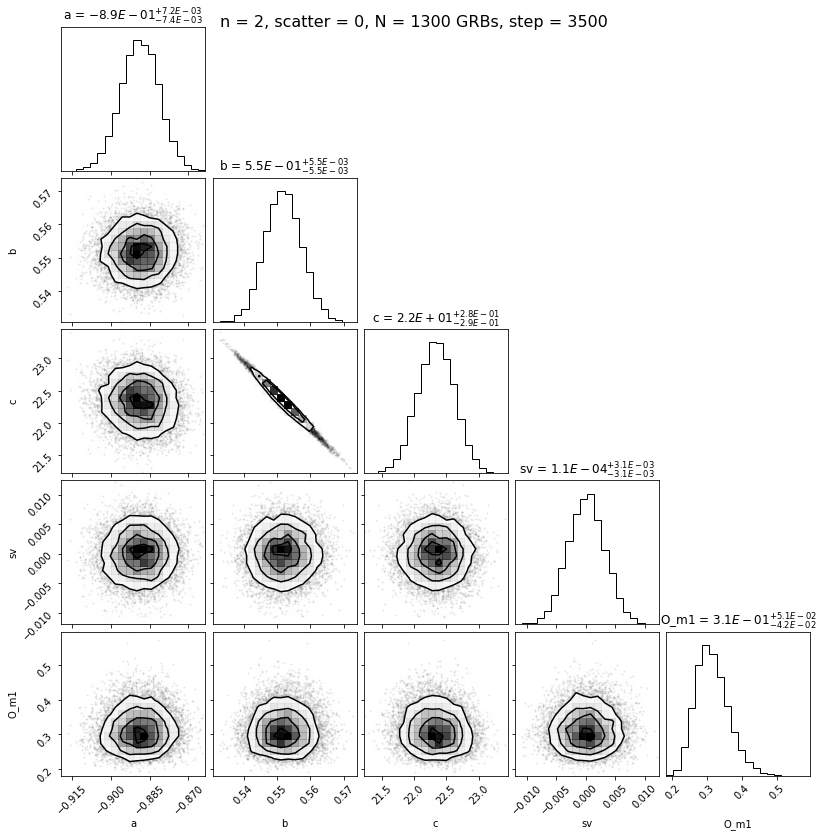}
    \includegraphics[scale=0.35]{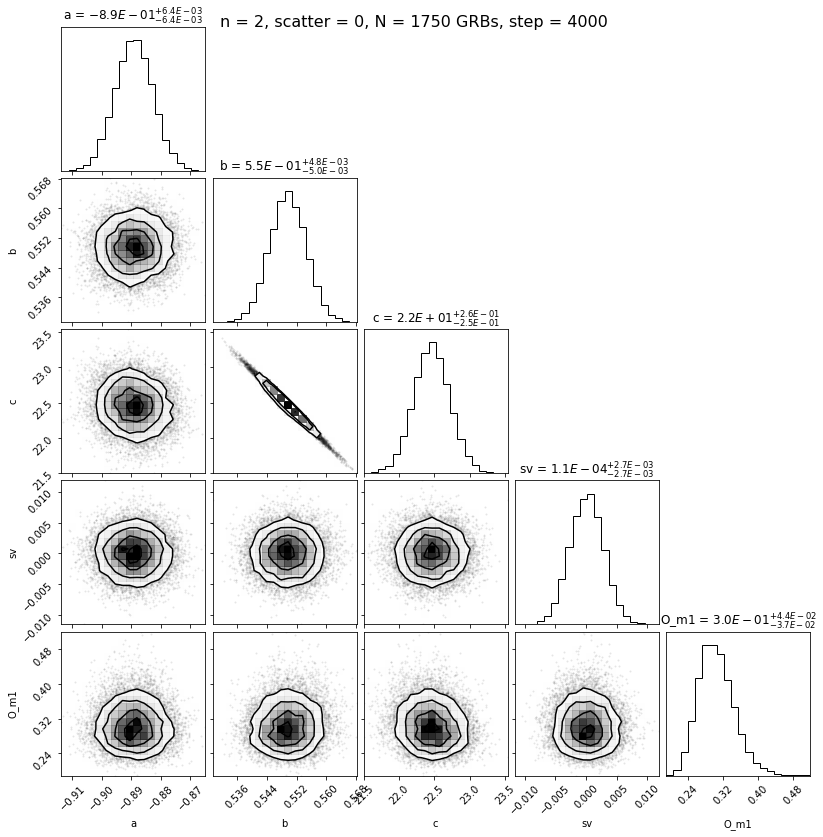}
    \caption{The posterior distribution for the 1300 (upper panel) and 1750 (lower panel) simulated GRBs with $n=2$, starting from the platinum sample properties. The parameters $a,b,c,sv$, and $O\_m1$ are $\alpha,\beta,\gamma,\sigma_i$, and $\Omega_{M}$, respectively.}
    \label{fig:GRBsimulations}
\end{figure}

We discuss how many GRBs are required to achieve the same precision as SNe Ia on the $\Omega_M$ parameter \citep{Dainotti2022MNRAS}. For the GRBs to be used as standardizable candles, it is crucial to leverage the fundamental plane relation: we here stress that this relation has the smallest intrinsic scatter among the multidimensional GRB correlations that involve the afterglow emission, namely, $\sigma_i=0.18\pm0.07$. Both the platinum sample \citep{DainottiPlatinum} of 50 X-ray GRBs \citep{DainottiPlatinum} and the 45 optical GRBs fundamental plane sample \citep{Dainotti2022ApJS} are tested as cosmological tools in combination with SNe Ia. 
The platinum sample is an improvement of the golden sample because it considers only plateaus with no flares, and with a duration $> 500$ sec.
The results in \citep{Dainotti2022MNRAS} show that both lead to the same level of precision on the estimation of the $\Omega_M$ parameter, namely, $\Omega_M(PLAT+SNeIa)=0.299\pm0.009$ and $\Omega_M(OPT+SNeIa)=0.299\pm0.009$, respectively. The same applies when (1) the platinum and the optical sample are \textit{trimmed}, namely, are reduced to samples of 20 X-ray and 25 optical GRBs that are the closest to their fundamental plane fitting, and (2) the GRB parameters are corrected for astrophysical biases or $z$-evolution effects through the EP method. These results shows how the fundamental planes in both wavelength regimes are promising cosmological tools. The simulations of GRBs that follow the same properties of the 50 GRBs X-ray platinum sample and the 45 optical GRBs are performed with the \textit{emcee} package \citep{ForemanMackey}. According to the preliminary simulations, only 150 X-ray GRBs are needed to achieve a reasonable value of $\Omega_M=0.387\pm0.473$; nevertheless, this 1 $\sigma$ precision level is not satisfying since it is comparable with the central value. Thus, further hypotheses must be applied before the GRB simulations to achieve the same precision level of SNe Ia standalone samples. To this end, as the basis for comparison, the following precision level is considered: $\Delta \Omega_M=0.042$ from Betoule et al. 2014 (B14) \citep{Betoule2014}. The forthcoming research in the field of machine learning (ML) techniques will have a beneficial effect on the observation of GRBs. First, the light curve reconstruction approaches applied to GRBs \citep{LCR} can halve the errors on GRB parameters. If the errors halving ($n=2$) is taken into account, it is possible to obtain $\Omega_M=0.416\pm0.177$, thus reducing the uncertainty on $\Omega_M$ by 63 $\%$ with respect to the case of the unhalved errors ($n=1$). Second, the ML methods that can be used to infer the redshift of GRBs will double the sample of GRBs with $z$, allowing the cosmological analysis of a more extensive collection of events. In this perspective, a key role is also played by the forthcoming missions dedicated to the GRBs observations, like the space mission THESEUS with a tentative launch date in 2037 and an expected rate of observed GRBs between 300 and 700 per year \citep{Amati2018}.
In this work, we simulated 1300 and 1750 GRBs that mimic the platinum sample properties in the case of $n=2$. The results of the posterior distributions are reported in Figure \ref{fig:GRBsimulations}. For 1300 GRBs, $\Omega_M=0.310\pm0.046$, while for the 1750 GRBs, we find $\Omega_M=0.310\pm0.040$ (both with symmetrized uncertainties). In both cases, the precision on $\Omega_M$ is comparable with the level reached in B14. Considering these results and the ratio of X-ray GRBs with platinum sample properties over the total number of observed GRBs (50/1064), we expect to obtain a sample of GRBs with such characteristics by 2044 if we include the contribution of LCR techniques to halve the errors and the application of ML approaches to double the sample of GRBs with redshift. The X-ray platinum sample is not the only one that can be used as a standard candle sample. Indeed, through the use of the optical fundamental plane correlation with the aid of the light curve reconstruction approach applied to GRBs \citep{LCR} and a sample of GRBs doubled thanks to the ML techniques to infer the redshift values, we can reach the same precision of B14 in less than ten years from now ($\sim$2032) \citep{Dainotti2022MNRAS}.

\section{Summary and conclusions}
We used the platinum sample of GRBs \citep{DainottiPlatinum} as a cosmological tool. The fundamental plane correlation, which finds its interpretation in the magnetar model, shows a small intrinsic scatter $\sigma_i=0.18\pm0.07$ when fitted on the platinum sample of GRBs. Thus, the fundamental plan proves to be a reliable cosmological tool in the case of GRBs as standalone cosmological probes. Indeed, through the simulation of 1300 and 1750 GRBs, we can see how the same precision as the SNe Ia analyzed in \citep{Betoule2014} can be reached in less than a decade from today. The forthcoming use of GRBs as standard candles will allow us to cast more light on the open problems of modern cosmology, in particular, the Hubble constant ($H_0$) tension for what concerns the standard $\Lambda$CDM cosmology \citep{Dainotti2021SNe,Dainotti2022SNe,Bargiacchi2023,Dainotti2023SNeIa,Lenart2023}.

\bigskip
\bigskip
\noindent {\bf DISCUSSION}

\bigskip
\noindent {\bf ANDREA ROSSI:} Why in the platinum sample you removed the GRBs with plateau duration $<500\,s$?\\
{\bf Answer:} The removal of the GRBs with plateau $<500\,s$ allows to consider all cases with plateus well-defined. In some cases infact some large time gaps happen in the lightcurve at that time and small plateau duration are not well-sampled. This then may lead to a non-optimal estimation of the plateau features, namely, the end-of-plateau time and luminosity.

\end{document}